\documentstyle[twocolumn,pra,aps,epsfig]{revtex}
\def\be{\begin{equation}}
\def\ee{\end{equation}}
\def\bea{\begin{eqnarray}}
\def\eea{\end{eqnarray}}
\begin{document}
\draft
%
%
\title{Quantum gates with neutral atoms: Controlling collisional interactions in time dependent traps}
\author{T. Calarco$^{1,2}$, E. A. Hinds$^3$, D. Jaksch$^1$, J. Schmiedmayer$^4$, J.I. Cirac$^1$, 
and P. Zoller$^1$}
\address{$^1$Institut f\"ur Theoretische Physik, Universit\"at Innsbruck,
Technikerstra\ss e 25/2, A--6020 Innsbruck, Austria\\
$^2$ECT*, European Centre for Theoretical Studies in Nuclear Physics
and Related Areas \\
Villa Tambosi, Strada delle Tabarelle 286, I--38050 Villazzano (Trento), Italy\\
$^3$ Sussex Centre for Optical and Atomic Physics, University of Sussex, Brighton, 
BN1 9QH, United Kingdom\\
$^4$ Institut f\"ur Experimentalphysik, Universit\"at Innsbruck, Technikerstra\ss e 25/4, 
A--6020 Innsbruck, Austria}
\date{\today}
\maketitle
%
%
\begin{abstract}
We theoretically study specific schemes for performing a fundamental two--qubit quantum gate 
via controlled atomic collisions by switching microscopic potentials. 
In particular we calculate the fidelity of a gate operation for a configuration where a potential 
barrier between two atoms is instantaneously removed and restored after a certain time.
Possible implementations could be based on microtraps created by magnetic and electric fields,
or potentials induced by laser light.
\end{abstract}
%
%
%
%

\section{Introduction}

The creation and manipulation of many--particle entangled states offers 
new perspectives for the investigation of fundamental
questions of quantum mechanics, and is the basis of applications such as
quantum information processing.
Several proposals to implement quantum logic \cite{QECC} have been made including 
ion-traps \cite{ions}, cavity QED and photons \cite{QED},
and molecules in the context of NMR \cite {NMR}. 
Very recently, we have identified a new
way of entangling neutral atoms by using {\it cold controlled collisions} \cite{jaksch99}
(see also \cite{brennen}).
Neutral atoms are good candidates for quantum information processing, since they suffer a 
comparatively weak dissipative coupling to the environment. Techniques
to cool and trap atoms by means of magnetic and optical potentials have been developed in
the context of laser cooling and trapping, and Bose--Einstein condensation (BEC) \cite{BEC}. 
In particular the ongoing development of magnetic microtraps \cite{microtraps}
offers an interesting new perspective
for storing and manipulating arrays of atoms \cite{hinds12,Schmiedmayer} 
and possible applications in quantum information \cite{hinds3}.

\begin{figure}[tbp]
\begin{center}
\epsfig{file=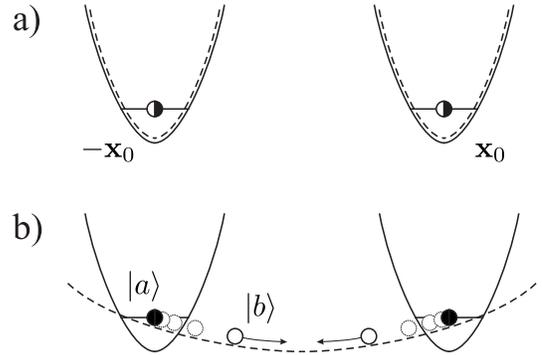,width=7cm}
\vspace*{0.5truecm}
\caption{Configuration at times $t<0$, $t>\tau$ (a) and during the gate operation 
(b). The solid (dashed) curves show the potentials for particles in the internal
state $|a\rangle$ ($|b\rangle$). 
\label{switching}}
\end{center}
\end{figure}

Motivated by these new experimental possibilities we will study in this paper specific configurations 
of atoms stored in time dependent microtraps. We will assume that two internal states of the
atoms $|a\rangle$ and $|b\rangle$ represent the logical states $|0\rangle$ and $|1\rangle$ of
a qubit, respectively.
The aim is to implement a fundamental two--qubit quantum gate between two atoms with 
the truth table 
\bea
|0\rangle|0\rangle&\rightarrow&|0\rangle|0\rangle,\nonumber\\
|0\rangle|1\rangle&\rightarrow&|0\rangle|1\rangle,\nonumber\\
|1\rangle|0\rangle&\rightarrow&|1\rangle|0\rangle,\nonumber\\
|1\rangle|1\rangle&\rightarrow&-|1\rangle|1\rangle,\label{phasegate}
\eea
by switching the trapping parameters. Eq.~(\ref{phasegate}) represents a so called phase gate.
To realize this transformation we will consider state selective switching of the trapping
potential such that the atoms pick up a phase due to collisional interaction \cite{schmiedphase} 
only if they are in state $|b\rangle$. This can be achieved by raising and lowering
a potential barrier between the two atoms as shown in Fig.~\ref{switching}.
According to Fig.~\ref{switching}a the potential is initially composed of 
two separated wells. Ideally the atoms have been cooled to the  vibrational ground states of the 
two wells. 
At time $t=0$ the shape of the trapping potential is changed for particles
in state $|b\rangle$ (dashed line in Fig.~\ref{switching}b) while the potentials for the atoms
in the state $|a\rangle$ remains unchanged (solid line in Fig.~\ref{switching}b). 
By removing the barrier the particles in state $|b\rangle$ start to oscillate and will collide.
\noindent
The ``cold'' collision represents a coherent interaction described by a pseudo--potential
with a strength proportional to the $s$--wave scattering length \cite{jaksch99}. This results in a phase 
shift of the wave function for both atoms in the internal state $|b\rangle$. The size of the phase 
shift can be controlled by the number of oscillations and the effective collisional interaction strength
(see Sec.~\ref{sham}).
As a last step the atoms have to be restored to the motional ground state of the trapping potential
of Fig.~\ref{switching}a. This whole process of switching the potentials can be performed
either as (i) switching the shape of the potential
{\it instantaneously} at times $t=0$ and $t=\tau$ where $\tau$ is a multiple of the oscillation 
period in the well of Fig.~\ref{switching}b (dashed line),
or, (ii) deforming the shape of the potential between Fig.~\ref{switching}a and b {\it adiabatically}.
The aim of the present paper is to investigate the gate dynamics for the scenario (i),
when the switching is instantaneous:
In particular we are interested in the required physical parameters and the corresponding fidelities
characterising the quality of the phase gate. We will also study the dependence of the fidelity
on the temperature of the atoms. The paper is organized as follows.
Section~\ref{sect:dynamics} describes the model and derives an expression for the 
collisional phase shift. In Sec.~\ref{sect:gate} we study the gate dynamics for the
case of instantaneous switching while in Sec.~\ref{sect:implementation}
we present numerical results for the fidelity.

\section{Model}
\label{sect:dynamics}

In the present section we will write down the Hamiltonian for two interacting particles
trapped in conservative time dependent potentials and derive an expression for the 
collisional phase shift.

\subsection{Hamiltonian}
\label{sham}

The dynamics of atoms in a time--varying, state--dependent trapping potential 
$V_\alpha({\bf x},t)$ (where $t$ is time and ${\bf x}\equiv(x,y,z)$
is the three-dimensional coordinate) can be described by the Hamiltonian operator
\bea
H&=&\sum_{\alpha\in\{a,b\}}\int d^3x\;\hat\Psi^\dagger_\alpha({\bf x})
\left[-\frac{\hbar^2}{2m}{\bf\nabla}^2+V_\alpha({\bf x},t)\right]
\hat\Psi_\alpha({\bf x})\nonumber\\
&&\mbox{}+\sum_{\alpha,\beta\in\{a,b\}}\frac 12\int d^3x\,d^3x'\,
\hat\Psi^\dagger_\alpha({\bf x})\hat\Psi^\dagger_\beta({\bf x}')\nonumber\\
&&\qquad\qquad\qquad\mbox{}\times U_{\alpha\beta}({\bf x},{\bf x}')
\hat\Psi_\beta({\bf x}')\hat\Psi_\alpha({\bf x}),\label{Hamil2q}
\eea
where $m$ is the mass of the atoms, $\hat\Psi_\alpha({\bf x})$ is a field 
operator for atoms in internal state
$|\alpha\rangle$, and $U_{\alpha\beta}({\bf x},{\bf x}')$ is the potential for
the interaction between two atoms in states $|\alpha\rangle$ and 
$|\beta\rangle$, where $\alpha,\beta\in\{a,b\}$. 
We take a trapping potential of the form
\be
\label{Pot3D}
V_\alpha({\bf x},t)=v_\alpha(x,t)+v_\perp(y)+v_\perp(z),
\ee
{\em i.e.} we assume the same shape 
along $y$ and $z$, which is independent of time and of the internal state. 

For cold atoms the dominant collisional interaction is the $s$--wave scattering term
described by a contact potential of the form
\be
\label{Ualphabeta}
U_{\alpha\beta}({\bf x},{\bf x}')=\frac{4 \pi a_s^{\alpha\beta} \hbar^2}{m} 
\;\delta^3 ({\bf x}-{\bf x}'),
\ee
where $a_s^{\alpha\beta}$ is the $s$--wave scattering length for the 
corresponding internal states. Note that for identical atoms in the same
internal state $s$--wave scattering is only possible for bosonic atoms
(cf.~the $b$--$b$ collision in Fig.~\ref{switching}b). 
We therefore require, in the following, the field operators $\hat\Psi_\alpha({\bf x})$ to describe 
bosonic atoms and to obey the usual bosonic commutation relations.

Furthermore, we
assume much stronger confinement along the $y$ and $z$ directions than in $x$, so that 
the probability of transverse excitations can be neglected. If each 
atom is initially in the ground state $|\psi_\perp\rangle$ of the transverse 
potentials, it will then remain in that state to a good approximation
and the corresponding degrees of freedom can be integrated out. 
In this case the dynamics becomes effectively one--dimensional
and is described by the Hamiltonian operator
\bea
H_x&=&\sum_{\alpha\in\{a,b\}}\int dx\;\hat\psi^\dagger_\alpha(x)
\left[-\frac{\hbar^2}{2m}\frac{d^2}{dx^2}+v_\alpha(x,t)\right]
\hat\psi_\alpha(x)\nonumber\\
&&\mbox{}+\sum_{\alpha,\beta\in\{a,b\}}\frac 12\int dx\,dx'\,
\hat\psi^\dagger_\alpha(x)\hat\psi^\dagger_\beta(x')\nonumber\\
&&\qquad\qquad\qquad\mbox{}\times u_{\alpha\beta}(x-x')
\hat\psi_\beta(x')\hat\psi_\alpha(x).\label{Ham1D}
\eea
Here $\hat\psi_\alpha(x)$ is the one--dimensional analogue of $\hat\Psi_\alpha({\bf x})$, and 
\bea
\label{ualphabeta}
u_{\alpha\beta}(x-x')&=&\int dy\,dy'\,dz\,dz'\, 
U_{\alpha\beta}({\bf x},{\bf x}')\nonumber\\
&&\quad\times
\left|\psi_\perp(y)\psi_\perp(y')\psi_\perp(z)\psi_\perp(z')\right|^2 \nonumber \\
&=&\frac{4 \pi a_s \hbar^2}{m} \delta(x-x') \left[\int dy |\psi_{\perp}(y)|^4 \right]^2,
\eea
is an effective interaction potential taking into account the
transverse confinement of the atoms. The $\psi_\perp$ are the ground--state 
wavefunctions in the transverse directions (having energy $\hbar\omega_\perp/2$ each). 
Their time evolution will just contribute an overall phase factor (with a phase
proportional to $\omega_\perp$), 
irrelevant for the quantities we are going to compute. We see that 
the effective interaction strength can be adjusted by changing
the trapping parameters.

Eq.~(\ref{Ham1D}) holds for an arbitrary number of atoms. 
We now consider the case of two bosonic atoms 1 and 2, with internal 
states $|a\rangle_{1,2}$ and $|b\rangle_{1,2}$. Their evolution is governed by 
the first--quantized Hamiltonian 
\be
{\cal H}=\sum_{\alpha,\beta\in\{a,b\}} {\cal H}_{\alpha\beta} \otimes 
|\alpha\rangle_1\langle \alpha| \otimes |\beta\rangle_2\langle \beta|,
\ee
where 
\begin{mathletters}
\bea
{\cal H}_{\alpha\beta}&\equiv &{\cal H}_{\alpha\beta}^0+u_{\alpha\beta},\\
{\cal H}^0_{\alpha\beta}&=&
{\cal H}_\alpha(p_1,x_1,t)+{\cal H}_\beta(p_2,x_2,t),\\
{\cal H}_\alpha(p_i,x_i,t)&=&\frac{p_i^2}{2m}+v_{\alpha}(x_i,t).
\eea
\end{mathletters}
Here $x_i$ and $p_i$ are the position and momentum operator for particle 
$i=1,2$ respectively.

\subsection{Phase shift due to interaction}
\label{sect:intphishift}

We call $|\psi_{\alpha\beta}^{(0)}(t)\rangle$ and 
$|\psi_{\alpha\beta}(t)\rangle$ the two--particle states
at time $t$ evolved from the same initial state $|\psi_{\alpha\beta}(0)\rangle$
in the absence and in the presence of interaction,
respectively:
\begin{mathletters}
\bea
i\hbar\partial_t|\psi^{(0)}_{\alpha\beta}(t)\rangle&=&
{\cal H}^0_{\alpha\beta}|\psi^{(0)}_{\alpha\beta}(t)\rangle,\label{Schroed0}\\
i\hbar\partial_t|\psi_{\alpha\beta}(t)\rangle&=&
{\cal H}_{\alpha\beta}|\psi_{\alpha\beta}(t)\rangle.\label{Schroed}
\eea
\label{Sch}
\end{mathletters}
We also define the overlaps
\begin{mathletters}
\bea
O_0(\psi_{\alpha\beta},t)&\equiv&
\langle\psi_{\alpha\beta}(t)|\psi_{\alpha\beta}^{(0)}(t)\rangle;\\
O(\psi_{\alpha\beta},t)&\equiv&
\langle\psi_{\alpha\beta}(t)|\psi_{\alpha\beta}(0)\rangle.
\eea
The condition that both atoms end up at time $t=\tau$ with  
the same spatial distribution they had at the beginning will not be exactly
fulfilled in realistic situations. However, in order for our scheme  
to work it is required that this is true at least approximately:
\end{mathletters}
\be
\label{Overlap1}
|O(\psi_{\alpha\beta},\tau)|\approx 1\qquad\forall\;\alpha,\beta,
\ee
{\em i.e.} the two--atom final state should differ 
from the initial one just by a phase factor 
$\Phi_{\alpha\beta}(\tau)\equiv\arg[O(\psi_{\alpha\beta},\tau)]$:
\be
\label{totphase}
|\psi_{\alpha\beta}(\tau)\rangle\approx e^{-i\Phi_{\alpha\beta}(\tau)}
|\psi_{\alpha\beta}(0)\rangle.
\ee
We also assume that the interaction between atoms does not induce any significant 
alteration in the shape of the wave functions, {\em i.e.}
\be
\label{Overlap01}
|O_0(\psi_{\alpha\beta},t)|\approx 1\qquad\forall\;\alpha,\beta,t.
\ee
Hence
\be
\label{collphase}
|\psi_{\alpha\beta}(t)\rangle\approx e^{-i\phi_{\alpha\beta}(t)}
|\psi_{\alpha\beta}^{(0)}(t)\rangle,
\ee
having defined the {\em collisional phase}
\be
\label{defcollphase}
\phi_{\alpha\beta}(t)\equiv\arg[O_0(\psi_{\alpha\beta},t)],
\ee
accounting for the 
contribution of the interaction to the total phase $\Phi_{\alpha\beta}(\tau)$.
The rest of the phase comes from the motion of the particles in the 
time--dependent trapping potential. From Eqs.~(\ref{Overlap1}) and 
(\ref{Overlap01}) it follows that
\be
\label{Overlap10}
|O(\psi^{(0)}_{\alpha\beta},\tau)|\approx 1\qquad\forall\;\alpha,\beta,
\ee
which implies, by analogy with Eq.~(\ref{totphase}),
\be
\label{kinphase}
|\psi_{\alpha\beta}^{(0)}(\tau)\rangle\approx
e^{-i[\phi_\alpha(\tau)+\phi_\beta(\tau)]}|\psi_{\alpha\beta}(0)\rangle.
\ee
Here the {\em kinematic phase} $\phi_\alpha(\tau)$ [$\phi_\beta(\tau)$] 
is defined 
as the phase that one atom would acquire after evolving for a time $\tau$ in 
the potential $v_\alpha$ [$v_\beta$] in the absence of the other particle.
By substituting Eq.~(\ref{kinphase}) into Eq.~(\ref{collphase}) evaluated at
$t=\tau$, and comparing it with Eq.~(\ref{totphase}), the collisional phase can 
be reexpressed as
\be
\label{phiarg}
\phi_{\alpha\beta}(\tau)\approx\Phi_{\alpha\beta}(\tau)-
[\phi_\alpha(\tau)+\phi_\beta(\tau)].
\ee 
By combining Eqs.~(\ref{Sch}), (\ref{Overlap01}) and (\ref{collphase}), we find
\be
\label{deltaE}
\hbar\partial_t\phi_{\alpha\beta}(t)\approx\langle\psi^{(0)}_{\alpha\beta}(t)|u_{\alpha\beta}
|\psi^{(0)}_{\alpha\beta}(t)\rangle\equiv\Delta E_{\alpha\beta}(t),
\ee
which is precisely the result one would expect from perturbation theory.
In order for Eq.~(\ref{Overlap01}) to hold, 
the time--dependent energy shift defined 
in Eq.~(\ref{deltaE}) has to satisfy the condition 
$\Delta E_{\alpha\beta}(t)\ll\hbar\omega$, 
with $\hbar \omega$ the first excitation energy of the system. 
Integration of Eq.~(\ref{deltaE}) gives a perturbative 
expression for the collisional phase:
\be
\label{phicol}
\phi_{\alpha\beta}(t)\approx\frac{1}{\hbar}\int_0^t dt'\,
\Delta E_{\alpha\beta}(t').
\ee

\section{Gate operation}
\label{sect:gate}

To proceed further, we have to specify the functional form of the potential 
$v_{\alpha}(x,t)$ in Eq.~(\ref{Pot3D}). 
The two atoms are initially trapped along $x$ in two separate harmonic
wells of frequency $\omega_0$, centered at $\pm x_0$. 
In order to simplify the analytic calculations, 
the confinement in the transverse directions is also assumed to be harmonic.
At $t=0$ the barrier between the wells is suddenly removed in a 
selective way for atoms in internal state $|b\rangle$:
an atom in state $\left| a\right\rangle $ feels no change, whereas one in
state $\left| b\right\rangle $ finds itself in a new harmonic
potential, centered on $x=0$ with frequency $\omega <\omega_0$. The
atoms are allowed to oscillate for some time, and then at $t=\tau$
the barrier is suddenly raised again to trap them at the original positions. 
During this 
process the atoms acquire a kinematic phase due to their oscillations within
the wells, and also --~if they collide~-- 
an interaction phase due to the collision.
Here we calculate these phases and consider the appropriate switching time $\tau$
for a quantum gate. In Sect. \ref{sect:implementation} we make a quantitative
estimate of the gate fidelity.

\subsection{\label{section:switching}Switching potential}

We take the potential in Eq.~(\ref{Pot3D}) to be explicitly
\begin{mathletters}
\bea
v_a(x,t)&=&\frac{m\omega_0^2}{2}\bigl[\theta(x)(x-x_0)^2+
\theta(-x)(x+x_0)^2\bigr];\label{defva}\\
v_b(x,t)&=&\left\{
\begin{array}{ll}
v_a(x,t)&t<0,\;t>\tau;\\
\displaystyle{\frac{m\omega^2}{2}x^2}&0\leq t\leq\tau;
\end{array}
\right.\label{defvb}\\
v_\perp(y)&=&\frac{m\omega_\perp^2}2y^2,
\eea
as shown in Fig.~\ref{switching}.
\end{mathletters}
As long as the single-well ground-state width $a_0=\sqrt{\hbar /m\omega_0}$
satisfies $a_0^2\ll x_0^2$ and there are no significant  
excitations to higher levels of $v_a(x,t)$, 
the actual behavior of that potential around the origin does not
really matter and we can use Eq.~(\ref{defva}) regardless of the experimental 
shape of the barrier around $x=0$.
The ground state wavefunctions $\psi_\pm(x)$ of the right and left well of the
potential $v_a(x,t)$ are given by
\bea
\psi_\pm(x)&=&
\left(\frac{m\omega_0}{2\pi \hbar }\right)^{\frac 14}
e^{-\frac{m\omega_0}{2\hbar }\left( x_0\mp x\right) ^2},
\eea
while the ground state wavefunction in the transverse directions is given by
\bea
\psi_\perp(y)&=&\left(\frac{m\omega_\perp}{2\pi \hbar }\right)^{\frac 14}
e^{-\frac{m\omega_\perp}{2\hbar }y^2}\label{psiperp}.
\eea 
By assumption the overlap between the two wavefunctions $\psi_+(x)$ and $\psi_-(x)$ is
negligible since the two particles are kept separated from each other in the
potential $v_a(x,t)$.
At $t=0$, the central barrier between the two wells is selectively switched off
for state $|b\rangle$. A particle in this state will start moving towards 
the other atom along $x$ and an interaction will take place. 
We shall separately study the evolution of the system  
at $t\geq 0$ for each combination of internal states $(\alpha,\beta)$.
For operation of the quantum gate analyzed here, it is important that $v_b(x,t)$
be accurately harmonic while $0\leq t\leq\tau$.

\subsection{Particles in the same internal state}

\subsubsection{Initial state}

If both particles are in the same internal state $\left| \alpha \right\rangle$, this 
factorizes from the motional degrees of freedom and the
initial state is
\be
\label{psisym}
|\psi_{\alpha\alpha}(0)\rangle=
\frac{|\psi_-\rangle|\psi_+\rangle+|\psi_+\rangle|\psi_-\rangle}{\sqrt{2}}
\otimes|\alpha\rangle|\alpha\rangle.
\ee
The calculation can be simplified by introducing the center of mass
(CM) and relative coordinates for the $x$--motion, thus rewriting
\bea
\label{psialphaalpha}
\psi_{\alpha\alpha}(x_1,x_2,0)&\equiv&\frac 1{\sqrt 2}
\left[\psi_-(x_1)\psi_+(x_2)+\psi_+(x_1)\psi_-(x_2)\right]\nonumber\\
&=&\psi_{\rm CM}(R,0)\psi_{\rm rel}(r,0),
\eea
where 
\begin{mathletters}
\bea
\psi_{\rm CM}(R,0) &=&\left( \frac{M\omega_0}{\pi \hbar }\right)
^{\frac 14}e^{-\frac{M\omega_0}{2\hbar }R^2}, \label{psi0CM}\\
\psi_{\rm rel}(r,0) &=&\left( \frac{\mu \omega_0}{4\pi \hbar }\right)^{\frac 14}
\sum_{\varsigma=-1,+1}e^{-\frac{\mu \omega_0}{2\hbar }
\left(2x_0+\varsigma r\right)^2}\label{psi0rel},
\eea
with $M=2m$, $\mu =m/2$, $R=\left( x_1+x_2\right) /2$, $r=x_2-x_1$.
\end{mathletters}

\subsubsection{Time evolution}
\label{sect:timevol}

For $t\leq 0$, the particles are stored in the displaced wells 
and no interaction takes place.
If both particles are in state $\left| a\right\rangle $,
the potential remains unchanged also for 
$t\geq 0$; there is no collision and thus the collisional phase 
$\phi_{aa}=0$. The state simply picks up the phase due to the free evolution: 
\be
\label{Evolaa}
|\psi_{\alpha\alpha}(t)\rangle=e^{-i\omega_0t}
|\psi_{\alpha\alpha}(0)\rangle.
\ee
We shall now consider the situation in which  
both particles are in state $\left| b\right\rangle $.
In this case, after the barrier
is switched off, the particles start oscillating in the harmonic trapping potential. 
In the absence of interaction they would come back to the initial state after an oscillation period 
$T_{\rm osc}=2\pi /\omega$,
having acquired a phase $4\pi\omega_\perp/\omega$ because of the transverse confining
potential. The interaction causes an additional phase to be accumulated by the wavefunction as
the number of oscillations increases, and a slight decrease in the oscillation
frequency, because the atoms acquire a small delay in their motion inside 
the trap as they come out from a collision. If the latter feature
is not too strong, by choosing a switching time $\tau\approx 2N\pi
/\omega $ it should be possible to get back the original state plus an
interaction phase, that is adjusted to $\pm \pi $ by a proper choice of the trap
parameters and of the number of collisions occurring during
the actual gate operation, \emph{i.e.} for $0<t<\tau$.
We shall therefore focus on the dynamics in this time interval. 

In the center of mass--relative coordinate system we get 
\be
{\cal H}_{bb}=\frac{P^2}{2M }+\frac{M \omega ^2}2R^2+
\frac{p^2}{2\mu }+\frac{\mu \omega ^2}2r^2+u_{bb}\left( r\right),
\label{hbb}
\ee
where $P=p_1+p_2$, $p=\left( p_1-p_2\right) /2$. If the interaction is neglected
we can solve the two--particle Schr{\"o}dinger equation for Hamiltonian
Eq.~(\ref{hbb}) analytically as shown in Appendix~\ref{app:sameanalyt}.
It can be seen from 
Eqs.~(\ref{EvolPsiCM}-\ref{OverPsiRel}) that the unperturbed
two-atom motion has a 
period of $T_{\rm osc}/2$ instead of $T_{\rm osc}$. 
This happens because the 
initial state, symmetric with respect to the origin, has 
nonzero projection only on the even eigenstates, having energies 
$(2n+1/2)\hbar\omega$: therefore, after a time $\pi/\omega$, each component of 
the wavefunction gets the same constant phase 
$\exp[i(2n+1/2)\pi]=\exp(i\pi/2)$. This has a simple physical interpretation:
if the atoms do not interact,
after half an oscillation period each particle is at its turning point,
coinciding with the other atom's starting location; so at that time
the two atoms have
interchanged their positions, but since they are indistinguishable this has to
be regarded as exactly the same motional state they had at the beginning 
(apart from a phase factor). 

When we take into account the interaction between particles, the center 
of mass motion is unaffected but the
relative motion can no longer be treated analytically.
The numerical method we use to carry out this calculation is 
outlined in Appendix~\ref{app:samenum}. It is, however, possible to take 
the interaction into account perturbatively as shown in the following section.

\subsubsection{Perturbative calculation of the phase shift}
\label{sect:phipert}

Eqs.~(\ref{ualphabeta}) and (\ref{psiperp}) combine to yield
\be
u_{\alpha\beta}(x_1-x_2)=2a_s^{\alpha\beta}\hbar\omega_\perp\delta(x_1-x_2).
\ee
When both particles are in state $|b\rangle$, the time--dependent 
energy shift defined in Eq.~(\ref{deltaE}) can be calculated analytically:
\bea
\Delta E_{bb}(t)&=&\int dR dr|\psi_{CM}(R,t)
\psi^{(0)}_{\rm rel}(r,t)|^2u_{bb}(r)\\
&=&a_s^{bb}\hbar\omega_\perp
\sqrt{\frac{8m\Omega (t)}{\pi \hbar }}
e^{-\frac{2m\omega_0}\hbar x_0^2
\left[1-\sin^2(\omega t)\frac{\omega_0\Omega(t)}{\omega^2}\right]}\nonumber,
\eea
where $\Omega(t)$ is defined in Eq.~(\ref{Omegat}).
The corresponding interaction-induced phase shift accumulated after an 
oscillation period is 
\be
\label{phiPert} 
\phi_{bb}(T_{\rm osc})\approx \frac{4a_s^{bb}\omega_\perp}{
\sqrt{x_0^2\omega^2-a_0^2\omega_0^2/4}},  
\ee
which has been evaluated by means of the well-known saddle--point approximation.

\subsection{Particles in different internal states}

\subsubsection{Initial state}

When the internal states of the atoms are different, they no longer factorize 
as in Eq.~(\ref{psisym}) and the initial state is given by
\be
\label{psialphabeta}
\left| \psi_{ab}(0)\right\rangle =\frac 1{\sqrt{2}}
\left[|\psi_-\rangle_1|\psi_+\rangle_2\otimes|a\rangle_1|b\rangle_2
+(1\leftrightarrow 2)\right],
\ee
where without loss of generality we assumed that the particle in the left
(right) well is in internal state $|a\rangle$  ($|b\rangle$).

\subsubsection{Time evolution}

The relevant quantities can again be expressed in terms of the 
projection of the 
evolved state on the initial one. By virtue of symmetry under particle
interchange, this turns out to be 
\be
O(\psi_{ab},t)=\langle\psi_-|\langle\psi_+|
e^{-\frac i\hbar{\cal H}_{ab}t}|\psi_-\rangle|\psi_+\rangle.
\ee
Therefore we can restrict our analysis, as in the previous case, to the
one--dimensional motion, starting from the non--symmetrized wavefunction
$\psi_-(x_1)\psi_+(x_2)$. 
The Hamiltonian for $0<t<\tau$ reads 
\bea
\label{H0ab}
{\cal H}_{ab}&=&\frac{p_1^2}{2m}+\frac{p_2^2}{2m}+\frac{m \omega_0^2}2\left(
x_1+x_0\right) ^2+\frac{m\omega ^2}2x_2^2 \nonumber\\
&&\quad\mbox{}+ u_{ab}(x_1-x_2) \nonumber \\
&=&\frac{P^2}{2M}+\frac{p^2}{2\mu}+
\frac m2(\omega^2-\omega_0^2)Rr+\frac m2\omega_0^2x_0^2
\left(1-\frac{\omega_0^2}{\tilde\omega^2}\right)\nonumber\\
&&\quad\mbox{}+\frac M2\tilde\omega^2
\left(R+\frac{\omega_0^2}{2\tilde\omega^2}x_0\right)^2+
\frac\mu 2\tilde\omega^2
\left(r-\frac{\omega_0^2}{\tilde\omega^2}x_0\right)^2  \nonumber \\
&&\quad \mbox{} +u_{ab}(r) 
\eea
where $\tilde{\omega}\equiv\sqrt{\left( \omega ^2+\omega_0^2\right) /2}$. 
Only the left well of $v_a(x_1,t)$ has been 
considered since the wavefunction remains
negligible in the region $x_1>0$ for $t>0$, as it is at $t=0$. It can
be seen from Eq.~(\ref{H0ab}) that the center of mass no longer decouples 
from the relative motion, unlike in the previous symmetrical case. 
A numerical calculation is needed to evaluate the phase shift $\phi_{ab}$.
This is done in Appendix~\ref{app:different}.

\subsection{Particles at finite temperature}

Up to now we have assumed the particles to be in a well known motional state.
In realistic experimental situations this may not be the case. The temperature
$T$ of the particles in the trap will be different from $0$ and thus the initial
state of the system with particles in internal states $\alpha$, $\beta$ is given 
by the density operator
\be
\label{rhoT}
\rho_{\alpha\beta}(T,t=0^-)\propto\mbox{e}^{- {\cal H}_{\alpha\beta}(0^-)/k_BT}.
\ee
This takes the average over different initial excited states, with a thermal 
probability distribution corresponding to $T$. As shown in Appendix
\ref{app:phiT} the collisional phase accumulated  
is independent of the shape of the wave function if the particles 
move at a constant velocity with respect to each other and the shape
of the one particle wavefunction does not change during the interaction. 
This is a good approximation for the interaction 
between particles in the same internal state $|b\rangle$.
The particles interact in the vicinity of the center of the well
where their velocity $v \approx x_0 \omega$ is almost constant and the 
shape of the one particle wavefunction does not change substantially as long 
as the conditions
\be
a_x \ll x_0, \qquad \mbox{and} \qquad a \ll x_0
\label{con}
\ee
hold, where $a$ is the width of the one particle wavefunction when the 
particles cross the center of the trap and $a_x=\sqrt{\hbar/m\omega}$. 
Therefore the collisional phase
$\phi_{bb}(T_{\rm osc})$ is almost independent of the temperature $T$
as long as mainly excitations fulfilling conditions Eq.~(\ref{con}) are
populated. Note that we are neglecting transverse excitations. 
If all three motional degrees of freedom are characterized by the same temperature
$T$, this is realistic as long as the condition $k_BT\ll\hbar\omega_\perp$ is satisfied. 
However, in principle it is also possible to cool the transverse motion separately, 
allowing a higher temperature along $x$.
Of course this would require that the rethermalization time is much larger than
the experimental time scale.
This lack of sensitivity to temperature applies quite generally, for example to atoms
interacting in an optical lattice as discussed in \cite{jaksch99}, provided that the
velocity at which the atoms are made to interact (in that case the velocity of
lattice movements) is kept constant during the interaction. 

\section{A physical implementation}
\label{sect:implementation}

We now consider the implementation of a switching potential by means of static
electric and magnetic trapping forces. We first discuss the possibility of obtaining 
the desired state dependence by means of devices which are experimentally 
available \cite{hinds12,hinds3}, when the present magnetic devices can be combined
with nanofabricated electrodes. Then we compute the performance of a
quantum gate for realistic trapping parameters.

\subsection{Microscopic electromagnetic trapping potential}
\label{sect:mirror}

The interaction between the magnetic dipole moment of an atom in some 
hyperfine state $|F,m_F\rangle$ and an external static magnetic field ${\bf B}$
entails an energy $U_{\rm magn}\approx g_F\mu_Bm_F|{\bf B}|$ 
depending on the atomic internal state via the quantum number $m_F$ (here
$\mu_B$ is the Bohr magneton and $g_F$ is the Land\'e factor). 
The Stark shift induced on an atom by an electric field 
${\bf E}$ gives an energy (independent on the hyperfine sublevel)
$U_{\rm el}\approx\frac 12\alpha_{\rm el} |{\bf E}|^2$, where
$\alpha_{\rm el}$ is the atomic polarizability. 
The interplay between these two effects
can be exploited in order to obtain a trapping potential whose shape depends 
on the internal state of the atoms.
As an example, we consider an atomic mirror like the one
recently realized \cite{hinds12} from a conventional video tape 
with sinusoidal magnetization ${\bf M}=(M_0\cos[k_Mx],0,0)$ along the $x$--axis. 
The period of the pattern, $2\pi/k_M$, can be as small as $1\mu$m with the system 
studied in \cite{hinds12}, or even close to 100 nm using existing magnetic storage technologies.
In order to get a microscopic trapping potential \cite{hinds3}, it is necessary to apply an 
external bias field ${\bf B}^{\rm ext}\equiv(0,B^{\rm ext}_y,B^{\rm ext}_z)$, 
oriented mainly along the $z$ axis, normal to the mirror's surface, 
and with a small component along $y$ in order to prevent trap losses 
due to spin flips occurring at magnetic field zeros.
In this case the magnetic trapping potential is
\begin{eqnarray}
V_{m_F}({\bf x})&=&g_F\mu_Bm_F\Big\{
B_0^2e^{-2k_Mz}\cos^2(k_Mx)+(B^{\rm ext}_y)^2\nonumber\\
&&\quad+\left[B_0e^{-k_Mz}\sin(k_Mx)+
B^{\rm ext}_z\right]^2\Big\}^\frac 12,\label{MagnPot}
\end{eqnarray}
where $B_0=\mu_0M_0(1-e^{-k_M\delta})/2$ and $\delta$ is the tape thickness.
The minima of $V_{m_F}$ form a periodic pattern
above the tape surface, at a height $z_0=\ln(\mu_0M_0/B_0)/k_M$
typically of the order of some fractions of $\mu$m. The
spacing between two nearest minima along $x$ is just the period of 
the magnetization, $2\pi/k_M$. 
With present-day technology, trapping frequencies can  range from
a few tens of kHz up to some MHz. Microscopic electrodes can be nanofabricated on the
mirror's surface \cite{Schmiedmayer}, thus allowing for the design of a 
potential with the characteristics described in Sect.~\ref{sect:gate}.

For the states $|a\rangle$ and $|b\rangle$ we choose the hyperfine
structure states $|a\rangle\equiv|F=1,m_F=-1\rangle$ and 
$|b\rangle\equiv|F=2,m_F=2\rangle$ of the $5S_{1/2}$ level of $^{87}$Rb, 
having scattering lengths $a_s^{bb}\approx a_s^{ab}\approx5.1$ nm.
Several schemes of loading atoms into the trap have been envisaged
(see for example \cite{hinds12,hinds3}). Most of them rely on an intermediate 
step, where atoms can be
trapped and cooled without coming in contact with the magnetic mirror. 
This pre--loading stage can be either a magnetic trap initially displaced from 
the surface, or a different kind of trap (for instance an 
evanescent wave mirror, where different internal states can be trapped by gravity
close to the surface \cite{GOST} before the atoms are put in the correct 
states for magnetic trapping), to be replaced by 
the electromagnetic microtrap with a gradual switch--on of the electric and bias
magnetic fields in the final stage of loading \cite{Schmiedmayer}. This could
also allow for implementing a controlled filling of the trap sites by 
adiabatically 
turning on the periodic potential, in a similar way to that discussed in
\cite{jaksch98}.

\subsection{Results}

\subsubsection{Time evolution during gate operation}
\label{sect:resultevol}

\begin{figure}
\epsfig{figure=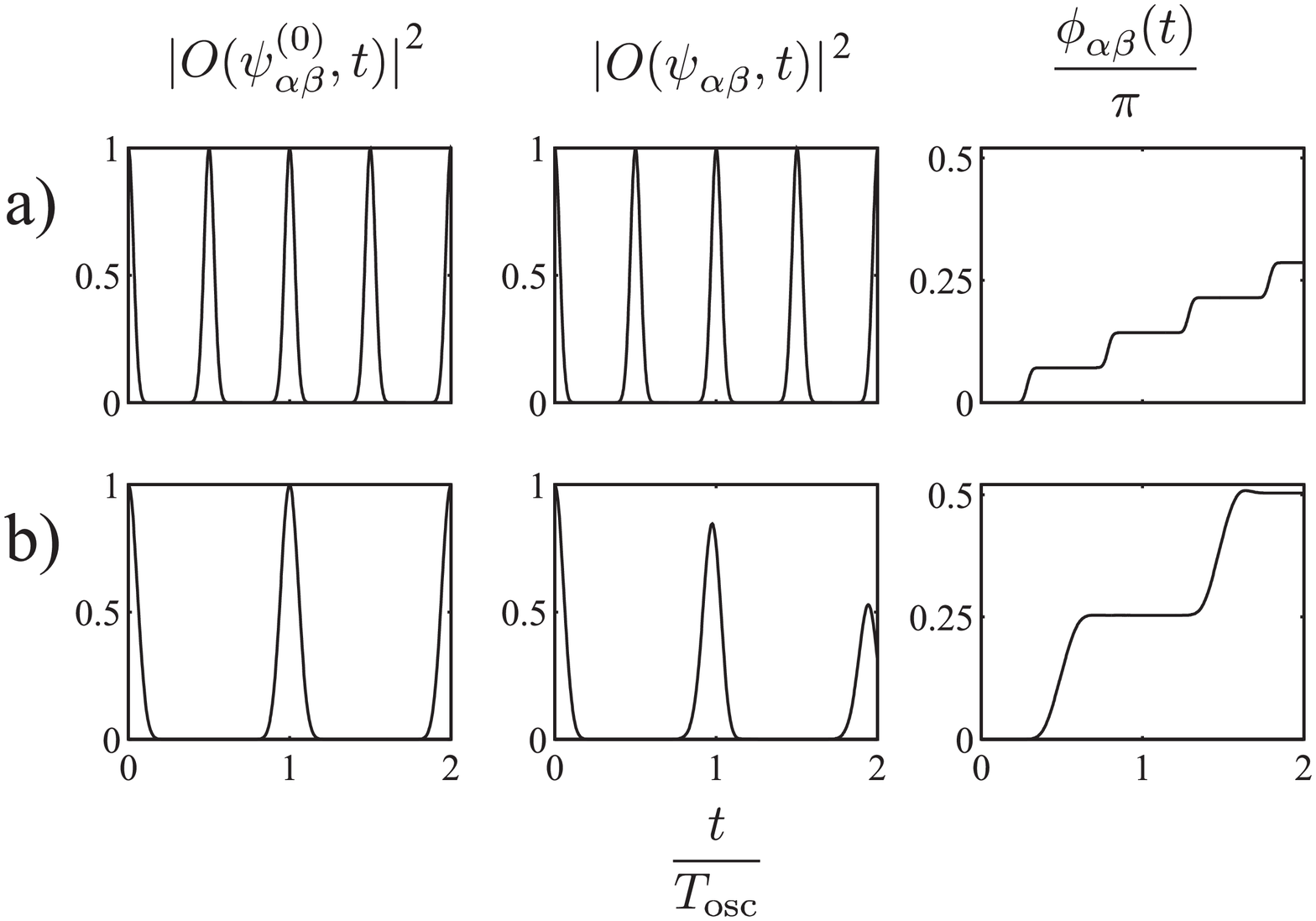,width=8truecm}
\caption{\label{Fig2}
Dynamics during gate operation: projection of the initial state on the state
evolved without (left) and with interaction (center); interaction--induced phase
shift (right). Results are shown for different combinations of internal states:
a) $\alpha=\beta=b$; b) $\alpha\not=\beta$.
We choose $\omega=2 \pi 17.23$ kHz and $\omega _\perp=2 \pi 150$ kHz,
corresponding to ground--state widths $a_x\approx 82$ nm, $a_\perp\approx 28$
nm, with the initial wells having frequency $\omega _0=2\omega $ and
displaced by $x_0=5a_x$. 
Time is in units of the oscillation period $T_{\rm osc}$.
}
\end{figure}

If both particles are in state $|a\rangle$, there is no interaction-induced 
phase shift, as expressed in Eq.~(\ref{Evolaa}).
The results for both particles in state $|b\rangle$ 
are shown in Fig.~\ref{Fig2}a, while those for differing internal states appear in 
Fig.~\ref{Fig2}b.
The harmonic potential ensures
that the system comes periodically back to its initial state. 
In the absence of interaction, the frequency of recurrencies is twice as
high for $|\psi_{bb}(t)\rangle$ as it is 
for $|\psi_{ab}(t)\rangle$, as already discussed at the end 
of Sect.~\ref{sect:timevol}. The interaction also makes the two cases substantially
different from each other. Its effect on the atomic motion is not dramatic if 
both particles are in state $|b\rangle$: actually, the oscillation period in 
the presence of interaction is increased by only
$\delta t\approx 1.4\times 10^{-3}T_{\rm osc}$ 
with the parameters used here. 
The collisional phase $\phi_{bb}$ increases in steps at the times 
$t_k\equiv(2k+1)T_{\rm osc}/4$,
when the atoms meet at the center of the well, and remains constant at
intermediate times while they are apart. 
Note that since the particles are indistinguishable the amplitude for the
particles to bounce back during the colllision does not harm the perfomance
of our scheme. The contributions of the reflected and the non-reflected part
to the wavefunction are indistinguishable.
What matters is whether of not the two--particle spatial distribution
approaches the initial one, and this is satisfied to a high accuracy in our
case.

The behavior is quite different if the atoms are in different internal states.
The phase shift increases in larger steps since the collision is close to
the turning point of the particle in state $|b\rangle$, near $x=x_0$. 
Here the velocity of the particle is much smaller than at the center of the
trap and thus the interaction time is longer, allowing a larger phase to accumulate. 
The collision also excites vibrations of the particle in state $|a\rangle$. 
The resulting loss of energy from the
particle in state $|b\rangle$ leads to a decreasing oscillation amplitude of that
particle, and the initial state is no longer recovered. This problem can be 
avoided if the potential minimum for state $%
\left| a\right\rangle $ is displaced along the transverse direction from the
one for state $\left| b\right\rangle $ by means of an additional electrostatic 
field \cite{hinds3}, so that the atoms interact if and only if they are both 
in state $\left| b\right\rangle$. 

\subsubsection{Gate fidelity at $T=0$}

Ideally, the scheme described above should realize the mapping 
\bea
|a\rangle|a\rangle &\rightarrow& 
  e^{-i2\phi_a} |a\rangle|a\rangle,\nonumber\\
|a\rangle|b\rangle &\rightarrow& 
  e^{-i(\phi_a+\phi_b+\phi_{ab})} |a\rangle|b\rangle,\nonumber\\
|b\rangle|a\rangle &\rightarrow& 
  e^{-i(\phi_b+\phi_a+\phi_{ab})} |b\rangle|a\rangle,\nonumber\\
|b\rangle|b\rangle&\rightarrow& 
  e^{-i(\phi_{bb}+2\phi_b)} |b\rangle|b\rangle,\label{mapping}
\eea
where $\phi_a$ and $\phi_b$ 
are the phases due to the time evolution without taking into account the 
interaction. 
We assume, as above, that the trapping potential is
designed to prevent the atoms interacting if they are in different
internal states. Therefore we set $\phi_{ab}=0$ in Eq.~(\ref{mapping}) and
consider only $\phi_{bb}$ in the following.
We use the minimum fidelity $F$ \cite{Schu} to characterize the quality of the 
gate. $F$ is defined as
\be
\label{Fid0}
F= \min_{\chi}{\bf\Bigl(}{\rm tr}_{\rm ext}\left\{\langle \tilde\chi| 
{\cal U}S \left[|\chi\rangle\langle\chi|\otimes \rho_0
\right]S^\dagger{\cal U}^\dagger|\tilde\chi\rangle\right\}{\bf\Bigr)},
\ee
where $|\chi\rangle$ is an arbitrary internal state of both atoms, and 
$|\tilde \chi\rangle$ is the state resulting from $|\chi\rangle$ using
the mapping (\ref{mapping}). The trace is taken over properly symmetrized
motional states, ${\cal U}$ is the evolution operator for the internal states
coupled to the external motion (including the collision), $S$ represents
symmetrization under particle interchange and $\rho_0$ 
is the density operator for the initial two--particle motional ground
state. A straightforward calculation gives
\be
F\!=\!\textstyle{\frac 12\frac{1-A^2-B^2[(1+A^2)B^2-4ABC+2C^2]
\cos^2(\phi_{bb})}
{(1+A)\{2+B[(1-A)B+2C]\cos(\phi_{bb})\}-B^2(B-C)^2\cos^2(\phi_{bb})}}
\ee
where $A=\bigl|O(\psi_{bb}^{(0)},\tau)\bigr|^\frac 12$,
$B=|O(\psi_{bb},\tau)|^\frac 12$, 
$C=\bigl|O_0(\psi_{bb},\tau)\bigr|^\frac 12$. With the parameters quoted
above, we obtain $F\approx 0.99$ either by choosing a gate operating time 
$\tau=7(T_{\rm osc}+\delta t)$, maximizing $B$, or 
$\tau=7T_{\rm osc}$, maximizing instead $A$.
We prefer this latter choice since, after a time $\tau=NT_{\rm osc}=2N\pi/\omega$, 
the $j^{\rm th}$ component of the $x$--wavefunction 
of an atom in state $|b\rangle$ in the basis of eigenstates 
of $v_b(0\leq t\leq\tau)$ gets a phase $2N(j+1/2)\pi$ (here $N=7$). 
This brings some simplifications: e.g.,
the kinematic phases can be written as 
\be
\phi_a=N\pi\frac{\omega_0+2\omega_\perp}\omega,\qquad
\phi_b=N\pi\frac{\omega+2\omega_\perp}\omega.
\ee
The general form of $\phi_b$ is much more complicated. 
Fig.~\ref{Fig3}a shows that after 7 complete oscillations
Eq.~(\ref{defcollphase}) yields a phase shift $\phi_{bb}(7T_{\rm osc})\approx\pi$
due to the interaction,
whereas the perturbative formula Eq.~(\ref{phiPert}) gives 
$7\phi_{bb}(T_{\rm osc})\approx 0.97\pi $. The figure also shows that the 
overlap $|O_0(\psi_{bb},t)\bigr|$ remains 
close to 1, satisfying Eq.~(\ref{Overlap01}).
The curve has local 
minima at the times $t_k$ defined in Sect.~\ref{sect:resultevol}, 
signalling that a collision is taking 
place, and shows a global decrease due to the accumulating  
delay of the interacting motion with respect to the noninteracting one. 
The fidelity turns out to be
\be
F=\frac 12\left\{1-\bigl|O_0(\psi_{bb},\tau)\bigr|
\cos\left[\phi_{bb}(\tau)\right]\right\}.
\ee

\subsubsection{Gate fidelity at $T\not=0$}

\begin{figure}
\epsfig{figure=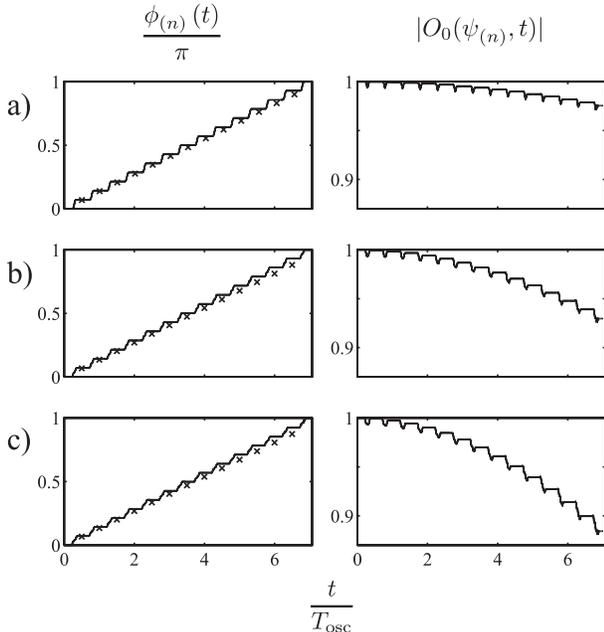,width=8truecm}
\caption{\label{Fig3}
Dynamics for both atoms in state $\left| b\right\rangle$, with relative--motion
excitations: a) $n=0$; b) $n=1$; c) $n=2$. 
On the left: interaction--induced phase shift; 
the crosses refer to the perturbative result from 
Eq.~(\protect\ref{phicol}), explicitly given by Eq.~(\protect\ref{phiPert})
for $n=0$, and evaluated numerically for $n>0$. On the right: 
projection of the evolved state on the corresponding state
evolved without interaction.
Trap parameters have the same values as in Fig.~\protect\ref{Fig2} and satisfy
Eq.~(\protect\ref{validveloc}) since $a_0\omega_0/(4x_0\omega)=0.07$ in this 
case.
}
\end{figure}

In order to compute the temperature dependence $F(T)$ of the fidelity, the 
density matrix for the motional degrees of freedom in Eq.~(\ref{Fid0}) has to 
be replaced by
\be
\rho_{\rm ext}(T)=\sum_{l,n}P_{ln}(T)|l\rangle_R\langle l|
\otimes|n\rangle_r\langle n|,
\ee
which coincides with $\rho_0$ at $T=0$.
Here we have introduced the eigenstates $|l\rangle_R$ for the center of mass and 
$|n\rangle_r$ for the relative motion.
The probabilities $P_{ln}(T)$ for occupation of the CM and relative motion 
excited states are calculated assuming for each atom a thermal distribution 
corresponding to temperature $T$, as expressed by Eq.~(\ref{rhoT}). 
We obtain
\be
\label{FidT}
F(T)=\frac 12\Bigl\{1-\sum_{l,n}P_{ln}(T)\bigl|O_0(\psi_{(n)},\tau)\bigr|
\cos\bigl[\phi_{(n)}(\tau)\bigr]\Bigr\},
\ee
where 
\bea
\psi_{(n)}(r)&=&\left( \frac{m\omega_0}{2\pi \hbar }\right)^{\frac 14}
\sum_{\varsigma=-1,+1}\frac{e^{-\frac{m\omega_0}{4\hbar}
\left(2x_0+\varsigma r\right)^2}}{\sqrt{n!2^{n+1}}}\nonumber\\
&&\qquad\qquad\mbox{}\times
H_n\left[\sqrt{\frac{m\omega_0}{2\hbar}}(2x_0+\varsigma r)\right].
\eea
In particular, $\psi_{(0)}\equiv\psi_{bb}$
and $\phi_{(0)}\equiv\phi_{bb}$.
The corresponding interaction--induced phase shifts 
$\phi_{(n)}(t)$ are shown in Fig.~\ref{Fig3}b,c.
The discrepancy between the interacting and the noninteracting motion increases
with $n$, but nevertheless the phase shift
$\phi_{(n)}$ remains still close to $\pi$ (Fig.~\ref{Fig3}b and c), 
as already discussed in Sect.~\ref{sect:phipert}. 
Consequently the fidelity is not rapidly suppressed with
temperature. 

For example, one might well be interested in the values of $F(T)$ for temperatures up to
$k_BT\approx\hbar\omega_0$. Let us therefore define $\gamma\equiv 
\exp(-\hbar\omega_0/k_BT)$ and neglect terms of $o(\gamma^7)$ in the
evaluation of Eq.~(\ref{FidT}) to obtain
\bea
F(T)&\approx&F(0)-\frac 12\sum_{n=1}^6\gamma^n\Bigl\{
\bigl|O_0(\psi_{(n)},\tau)\bigr|\cos\left[\phi_{(n)}(\tau)\right]\nonumber\\
&&\qquad\;\;\mbox{}-
\bigl|O_0(\psi_{(n-1)},\tau)\bigr|\cos\left[\phi_{(n-1)}(\tau)\right]\Bigr\}.
\eea
This still gives a high fidelity $F(T)\approx 0.96$ even at $k_BT=2\hbar\omega_0$, for which  
$\gamma^7\approx 0.03$. We note that, 
in order to reach such a high fidelity, the timing has to be quite precise, with a
resolution better than $10^{-3}T_{\rm osc}$ corresponding to tens of ns
in this case.

\section{Conclusions}
\label{sect:conclusions}

We have shown that entanglement among ultracold neutral atoms can be controlled 
by means of microscopic switching potentials. The fidelity for a
fundamental two--qubit quantum gate turns out to be quite robust with respect to
temperature: in fact, with the parameters quoted below Fig.~\ref{Fig2},
we find $F(T)\approx 0.96$ for $T\approx 3\mu$K in the $x$--motion, while
assuming ground--state cooling in the transverse directions. 
We find a gate operation time of $\tau\approx 0.4$ms, over which 
coherence can probably be preserved with presently available experimental systems.
Static microtraps based on available atomic mirrors \cite{hinds12,hinds3} provide 
a good opportunity for a first implementation of our scheme.
Here nanofabrication technologies allow steep potentials to be achieved with small 
charges and/or currents. Trapping fields can be controlled electronically in a
fast and accurate way \cite{Schmiedmayer}. 

Some problems remain to be addressed. To perform even a single gate
operation, the trap should be loaded with exactly one atom per well. 
Read-out should be done possibly without removing atoms from the trap. 
In order to build up more complex operations, gates should be arranged in a
periodic structure where coherent atom transport may take place between
different locations. This would permit gate operations either on one 
pair of atoms at a time, or on several pairs in parallel, a fact which could
be exploited for efficient implementation of quantum error correcting schemes
and fault--tolerant quantum computing \cite{JMO}. This will be the subject of
future work.

\acknowledgments 

We thank S.~A.~Gardiner for many useful discussions.
One of us (T.~C.) thanks M.~Traini and S.~Stringari for the kind hospitality at
the Physics Department of Trento University, and the ECT* 
for partial support during the completion of this work.
This work was supported in part by the \"Osterreichischer Fonds zur 
F\"orderung der wissenschaftlichen Forschung, 
the European Community under the TMR networks ERB-FMRX-CT96-0087 and Nanofab, the 
UK Engineering and Physical Sciences Research Council, and the 
Institute for Quantum Information GmbH.

\appendix

\section{Time evolution}

\subsection{Analytical calculation}
\label{app:sameanalyt}

If both particles are in state $|b\rangle$ we start from the Hamiltonian
Eq.~(\ref{hbb}), neglect the interaction term, and solve
the Schr\"odinger equation. We find  (omitting the internal state indices $bb$)
\be
\label{EvolPsiCM}
\psi_{\rm CM}(R,t)=\left[ \frac{M\Omega (t)}{\pi \hbar }\right]
^{\frac 14}e^{i\phi_{\rm CM}(R,t)-\frac{M\Omega (t)}{2\hbar }R^2},
\ee
where 
\bea
\Omega (t) &=&\frac{\omega ^2\omega_0}{\left[ \omega ^2\cos%
^2(\omega t)+\omega_0^2\sin^2(\omega t)\right] }, \label{Omegat}\\
\phi_{\rm CM}(R,t) &=&\frac{M\Omega (t)}{2\hbar }\frac{\omega
_0^2-\omega ^2}{\omega_0\omega }R^2\cos(\omega t)\sin%
(\omega t)-\frac{\omega t}2 \nonumber\\
\label{phiCM} &&
-\frac 12{\rm arctan}\left[ \frac{\left( \omega_0-\omega \right) 
\cos(\omega t)\sin(\omega t)}{\omega \cos^2(\omega
t)+\omega_0\sin^2(\omega t)}\right] . \! 
\eea
From Eqs.~(\ref{psi0CM}) and (\ref{EvolPsiCM}) it follows
\begin{equation}
\label{OverPsiCM}
\left|O\left(\psi _{\rm CM},t\right)\right| ^2=
\left[ 1+\frac{\left(\omega_0^2-\omega^2\right)^2}
{4\omega_0^2\omega^2}\sin^2\left( \omega t\right)\right] ^{-\frac 12}.
\end{equation}
If the particles did not interact, the relative motion would be
\bea
\label{EvolPsiRel}
\psi_{\rm rel}^{(0)}(r,t)&=&\sqrt{\frac{\mu\Omega(t)}{4\pi\hbar}}
\left(e^{i\phi_{\rm rel}(-r,t)-\frac{\mu \Omega (t)}{2\hbar }%
\left[ r+2x_0\cos(\omega t)\right] ^2}\right.\nonumber\\ 
&&\qquad\;
\left.\mbox{}+e^{i\phi_{\rm rel}(r,t)-
\frac{\mu \Omega (t)}{2\hbar }\left[ r-2x_0\cos(\omega t)\right] ^2}\right),
\eea
where 
\bea
\phi_{\rm rel}(r,t)&=&\textstyle{-\frac{\omega t}2
-\frac 12{\scriptstyle\arctan}\left[ \frac{\left( \omega
_0-\omega \right) \cos(\omega t)\sin(\omega t)}{\omega 
\cos^2(\omega t)+\omega_0\sin^2(\omega t)}\right]}\\
&&\textstyle{+\frac{2\mu \Omega (t)}{\hbar \omega \omega_0}
{\scriptstyle\sin(\omega t)}
\left[ \bigl(\frac{\omega_0^2-\omega ^2}4{\scriptstyle r^2+
\omega_0^2x_0^2}\bigr){\scriptstyle \cos(\omega t)+\omega_0^2x_0r}\right] }.
\nonumber
\eea
The overlap between the states Eqs.~(\ref{psi0rel}) and (\ref{EvolPsiRel}) is 
\bea
\label{OverPsiRel}
\left|O\bigl(\psi_{\rm rel}^{(0)},t\bigr)\right|^2
&=&\Biggl( e^{-\frac{8m\omega_0\omega ^2x_0^2\cos^2(\omega t)}
{\hbar\omega_{+}^2(t)}}
+e^{-\frac{8m\omega_0\omega ^2x_0^2\sin^2(\omega t)}
{\hbar\omega_{-}^2(t)}}\nonumber\\
&&\quad\mbox{}+\frac{2\cos\left[ \frac{4m\omega }\hbar%
\frac{\omega_0^2(\omega_0^2+\omega ^2)}{\omega_{+}^2(t)\omega_{-}^2(t)}
{\scriptstyle x_0^2}\right]}{e^{\frac{4m\omega_0}\hbar \left[ \frac{\cos%
^2(\omega t)}{\omega_{+}^2(t)}+
\frac{\sin^2(\omega t)}{\omega_{-}^2(t)}\right] \omega ^2x_0^2}}\Biggm)
\nonumber\\
&&\mbox{}\times\left[\textstyle{ 1+\frac{(\omega_0^2-\omega^2)^2}
{4\omega_0^2\omega^2}\sin^2\left( \omega t\right)}\right] ^{-\frac 12}
\eea
\noindent with $\omega_{\pm }(t)=\sqrt{\omega ^2+\omega_0^2\pm \left( \omega
^2-\omega_0^2\right) \cos\left( \omega t\right) }$. 

This result for the relative motion
should be compared to the actual evolution in the presence
of interaction, which cannot be computed analytically. 
If the particles are in different internal states we also have to
resort to numerical methods.

\subsection{Numerical calculation}

\subsubsection{Particles in the same internal state}
\label{app:samenum}

We write the state vector as a sum over the
eigenstates $\left| n\right\rangle $ of a harmonic oscillator of mass $\mu $ and
frequency $\omega$, 
\be
\left| \psi_{\rm rel}(t)\right\rangle =\sum_ne^{-i\left( n+1/2\right)
\omega t}c_n(t)\left| n\right\rangle
\ee
and approximate the potential by a truncated sum
\bea
\delta(r)&\approx &\sum_{k,l}^{N_{\rm max}}
\left| k\right\rangle \left\langle
k\right| \delta (r)\left| l\right\rangle \left\langle l\right|\nonumber\\
&=&\sum_{k,l}^{N_{\rm max}}\psi_k^{*}(0)\psi_l(0)|k\rangle \langle l| ,
\eea
where $\psi_n(x)=\left\langle x|n\right\rangle $. 
We have checked that the final result is independent of $N_{\rm max}$,
with $N_{\rm max}$ of the order of some tens. The Schr\"{o}dinger
equation for $\left| \psi_{\rm rel}(t)\right\rangle $ gives 
\be
\dot{c}_n(t)=-i 2a_s^{bb} \omega_\perp
\psi_n^{*}(0)\sum_{l=0}^{N_{\rm max}}\psi_l(0)e^{i(n-l)\omega t}c_l(t),
\ee
which we solve numerically for $c_n(t)$ with $0 \leq n \leq N_{\rm max}$.
The initial conditions, from Eq.~(\ref{psi0rel}), read 
\bea
c_n(0) &=&\frac{e^{-\frac{m\omega_0\omega }{\hbar (\omega_0+\omega )}%
x_0^2}}{\sqrt{n!2^n}}%
\frac{(\omega_0\omega)^{\frac 14}}{\sqrt{\omega_0+\omega}}
\left( \frac{\omega_0-\omega }{\omega_0+\omega }\right) ^{\frac n2}\\
&&\textstyle{\times\left[ H_n\Bigl(\sqrt{\frac{2m\omega\omega_0^2x_0^2}
{\hbar (\omega_0^2-\omega ^2)}}\Bigm)
+H_n\Bigl( \sqrt{\frac{2m\omega \omega_0^2x_0^2}{%
\hbar (\omega^2-\omega_0 ^2)}}\Bigm) \right] }. \nonumber
\eea

\subsubsection{Particles in different internal states}
\label{app:different}

In order to solve the Schr\"{o}dinger equation for the Hamiltonian Eq.~(\ref{H0ab})
we decompose the state vector
\be
\left| \psi_{ab} (t)\right\rangle =\sum_{j,k}e^{-i\left( j+k+1\right) \tilde{%
\omega}t}c_{jk}(t)| \tilde{j}\rangle_R| \tilde{k}\rangle_r,
\ee
(where now $\tilde{\psi}_j(x)=\langle x|\tilde{j}\rangle $ are
the eigenfunctions of a harmonic oscillator with frequency 
$\tilde{\omega}$ and mass $m$) and obtain for the coefficients
\bea
\dot{c}_{jk}(t) \!&=&i\frac{\tilde{\omega}(\omega_0^2-\omega ^2)}{2\left(
\omega_0^2+\omega^2\right) }\Big\{ c_{j+1,k+1}(t)e^{-i2\tilde{\omega}t}%
\sqrt{\left( j+1\right) \left( k+1\right) }\nonumber\\
&&\mbox{}+c_{j-1,k+1}(t)\sqrt{j\left(
k+1\right) }+c_{j-1,k-1}(t)e^{i2\tilde{\omega}t}\sqrt{jk} \nonumber \\
&&\mbox{}+c_{j+1,k-1}(t)\sqrt{\left( j+1\right) k}+
\frac{m\tilde{\omega}}\hbar \frac{\omega ^2-\omega_0^2}{\omega_0^2} \xi
^2c_{jk}(t) \nonumber \\
&&\mbox{}+\sqrt{\frac{2m\tilde{\omega}}\hbar }\xi
\Big[ e^{i\tilde{\omega}t}\big(c_{j-1,k}(t)\sqrt{j}
-c_{j,k-1}(t)\sqrt{k}\big)\nonumber\\
&& \mbox{}+e^{-i\tilde{\omega}t}\big(c_{j+1,k}(t)\sqrt{j+1}
-c_{j,k+1}(t)\sqrt{k+1}\big)\Big]\Big\} \nonumber\\
&& \mbox{}-i \sqrt{2} a_s^{ab} \omega_\perp
\tilde{\psi}_k^{*}
(-\xi )\sum_l\tilde{\psi}_l^{*}(-\xi )e^{i(k-l)\tilde{\omega}t}c_{jl}(t)
\nonumber\\
&&
\eea
where $\xi =x_0\omega_0^2/\sqrt{2}\tilde\omega ^2$. 

This can again be solved numerically, starting from the initial conditions,
derived from Eqs. (\ref{psi0CM}), (\ref{psi0rel}), 
\bea
c_{jk}(0) &=&\frac{e^{-\frac{m\omega_0\omega (\xi ^2+x_0^2-\sqrt{2}x_0\xi )}
{\hbar (\omega_0+\omega )}}}{\sqrt{j!k!2^{j+k-2}}}
\frac{\sqrt{\omega_0\omega }}{\omega_0+\omega }
\Bigl( \frac{\omega_0-\omega }{\omega_0+\omega }\Bigr) ^{\frac{j+k}2}\!\\
&&\textstyle{\times\left[ H_j\Bigl(\sqrt{
\frac{m\omega\omega_0^2(\sqrt{2}x_0-\xi)^2}{\hbar(\omega_0^2-\omega^2)}}\Bigr)
+H_k\Bigl( \sqrt{\frac{m\omega \omega_0^2\xi^2}{%
\hbar (\omega_0^2-\omega^2)}}\Bigr) \right] }. \nonumber
\eea

\section{Interaction phase shift for excited initial states}
\label{app:phiT}

Let us consider two bosonic atoms in the same internal state $|\alpha\rangle$, 
but in two different single--particle motional states $|\varphi_-\rangle$ and 
$|\varphi_+\rangle$ with vanishing overlap. 
The initial motional state has the form
\be
\label{psiexc}
|\varphi(0)\rangle=
\frac{|\varphi_-\rangle|\varphi_+\rangle+
|\varphi_+\rangle|\varphi_-\rangle}{\sqrt{2}}.
\ee
We assume that: (i) the particles
move against each other, come in contact during a certain time interval 
$[t_i,t_f]$ and then separate again; (ii) the velocity of each particle and
the shape of its wavefunction do not vary during the interaction.
Thus for $t_i\leq t\leq t_f$ we write:
\begin{mathletters}
\bea
\langle x_1|\varphi_-(t)\rangle &=&\varphi'(x_1-vt),\label{varphi1}\\
\langle x_2|\varphi_+(t)\rangle&=&\varphi''(x_2+vt),\label{varphi2}
\eea
where $v$ is a positive constant. It follows
\end{mathletters}
\bea
\phi_{\alpha\alpha}(T_{\rm osc})&\approx&\frac 1\hbar\int_{t_i}^{t_f}dt\;
\langle\varphi(t)|u_{\alpha\alpha}(x_1,x_2)|\varphi(t)\rangle\nonumber\\
&=&4a_s^{\alpha\alpha}\omega_\perp\int_{t_i}^{t_f}dt\int_{-\infty}^{+\infty}dx_1
\nonumber\\&&\qquad\mbox{}\times
\left|\varphi'(x_1-vt)\right|^2\left|\varphi''(x_1+vt)\right|^2\nonumber\\
&\approx &\frac{2a_s^{bb}\omega_\perp}{v}\int_{-\infty}^{+\infty} dx \, dy \,
\left|\varphi'(x)\right|^2\left|\varphi''(y)\right|^2\nonumber\\
&=&\frac{2a_s^{bb}\omega_\perp}{v}\label{phiexc},
\eea
where a change of variables $x=x_1-vt$, $y=x_1+vt$ has been introduced, and the
limits of integration in $t$ have been extended to $\pm\infty$ since the
single--particle wavefunctions Eqs.~(\ref{varphi1})-(\ref{varphi2}) overlap just
for a finite time. The result turns out to be independent of the initial state.
We can compare it to Eq.~(\ref{phiPert}), which was 
obtained in the harmonic potential Eq.~(\ref{defvb}) starting from the
single--particle states $|\psi_\pm\rangle$ instead of $|\varphi_\pm\rangle$. 
In this case
\be
v\equiv\left|\partial_t\langle\psi_\pm|e^{\frac i\hbar{\cal H}_bt}x\,
e^{-\frac i\hbar{\cal H}_bt}|\psi_\pm\rangle\big|_{t=t_k}\right|=x_0\omega,
\ee
and the atoms collide twice during one oscillation period. Therefore
the collisional phase Eq.~(\ref{phiPert}) should be twice as big as
Eq.~(\ref{phiexc}). This is true provided that 
the maximum velocity for the atomic motion in the well 
$v_b(x,0\leq t\leq\tau)$ is large with respect to the analogous quantity 
for the ground--state motion in the wells 
$v_a(x,t)$, {\em i.e.} if
\be
\label{validveloc}
x_0\omega\gg a_0\omega_0/4.
\ee

%
%

\end{document}